\documentclass[12pt]{article}

\usepackage{amsmath}
\usepackage{amssymb}

\newcommand{\1}{\,\mathbb{I}}
\newcommand{\bra}[1]{\left\langle{}#1\right|}
\newcommand{\cbn}{\mathbf{S}}

\newcommand{\dfd}[1]{\emph{#1}}
\newcommand{\dst}{\mu}
\newcommand{\ee}{\mathbf{e}}

\newcommand{\entrp}{S}
\newcommand{\fbs}{\mathcal{F}}
\newcommand{\gbs}{\mathcal{G}}
\newcommand{\hh}{\mathcal{H}}
\newcommand{\ket}[1]{\left| #1\right\rangle}

\newcommand{\lmu}{\lambda}
\newcommand{\lth}{d}
\newcommand{\mesb}{\,{\rm d} \phi}
\newcommand{\pdr}[2]{\frac{\partial\,#1}{\partial\,#2}}
\newcommand{\prxp}[1]{{#1}\otimes{#1}'}
\newcommand{\rme}{{\rm e}}
\newcommand{\rrh}{\rho}
\newcommand{\tpar}{\beta}

\newcommand{\thb}{m}
\newcommand{\tsth}{M}

\DeclareMathOperator{\trc}{Tr}

\newcommand{\bracket}[3]{\bra{#1}#2\ket{#3}}
\newcommand{\braket}[2]{\left\langle{}#1\right|\left.#2\right\rangle}
\newcommand{\ketbra}[2]{\ket{#1}\!\!\bra{#2}}
\newcommand{\tsthp}{\mathbf{\tsth}}
\newcommand{\tldh}{\tilde{\tsth}}

\newcommand{\prjo}[1]{\ketbra{#1}{#1}}

\title{How `hot' are mixed quantum states?}

\author{George Parfionov\thanks{Friedmann Laboratory For Theoretical Physics,
Department of Mathematics, SPb EF University, Griboyedova 30--32,
191023 St.Petersburg, Russia} and Rom\`an R.
Zapatrin\thanks{Department of Information Science, The State Russian
Museum, In\.zenernaya 4, 191186 St.Petersburg, Russia; e-mail:
zapatrin@rusmuseum.ru (corresponding author) }}

\date{}

\begin{document}

\maketitle

\begin{abstract}
Given a mixed quantum state $\rho$ of a qudit, we consider any
observable $M$ as a kind of `thermometer' in the following sense.
Given a source which emits pure states with these or those
distributions, we select such distributions that the appropriate
average value of the observable $M$ is equal to the average
Tr$M\rho$ of $M$ in the stare $\rho$. Among those distributions we
find the most typical one, namely, having the highest differential
entropy. We call this distribution conditional Gibbs ensemble as it
turns out to be a Gibbs distribution characterized by a
temperature-like parameter $\beta$. The expressions establishing the
liaisons between the density operator $\rho$ and its temperature
parameter $\beta$ are provided. Within this approach, the uniform
mixed state has the highest `temperature', which tends to zero as
the state in question approaches to a pure state.
\end{abstract}

\noindent Keywords: \emph{quantum ensembles; differential entropy;
Gibbs distribution}

\section*{Introduction}    

It is a notorious property of quantum systems that their mixed
states can be prepared using non-equivalent ensembles. For instance,
having a qubit, if you mix two pure states $(0,1)$ and $(1,0)$ with
proportion $\frac{1}{2}:\frac{1}{2}$, or three states
$\frac{1}{\sqrt{281}}\left(9,10\sqrt{2}{\rm i}\right)$,
$\frac{1}{\sqrt{194}}\left(12,5\sqrt{2}{\rm i}\right)$ and
$\frac{1}{\sqrt{17}}\left(3{\rm i},2\sqrt{2}\right)$ in proportion
$\frac{281}{900}:\frac{97}{450}:\frac{17}{36}$, you get the same
mixed quantum state with the density matrix \cite{hujowoo}:
\[
\rrh=\frac{1}{2} \left(
\begin{array}{cc}
1 & 0 \\ 0 & 1
\end{array}
\right)
\]

Now consider a continuous probability distribution on the set of all
pure states (=one-dimensional projectors $\prjo{\phi}$), denote its
density by $\dst(\phi)$. We may view this as a source of particles
which are emitted according to the probabilistic distribution
$\dst(\phi)$. To be rigorous, note that the characteristic feature
of $\dst$ is
\begin{equation}\label{enormalization}
  \int_{\cbn}\;
  \dst(\phi)
  \,\mesb
  \;=\;
  1
\end{equation}
where $\cbn$ is the set of all pure states and $\mesb$ is the
unitary invariant measure on pure states normalized to integrate to
unity. This measure $\mesb$ is unambiguously defined as $\cbn$ is a
connected compact manifold with a transitive action of the unitary
group $U(\lth)$.

Our next step is to fix an observable $\tsth$, associated with the
self-adjoint operator, denote it also by $\tsth$. For any pure state
$\prjo{\phi}$ the average value of $\tsth$ is defined equal to
$\bracket{\phi}{\tsth}{\phi}$, which is, in turn, a continuous
bounded function
\begin{equation}\label{emasfun}
\tldh(\phi)
\;=\;
\bracket{\phi}{\tsth}{\phi}
\end{equation}
on the manifold $\cbn$. Averaging the average \eqref{emasfun} with
respect to the distribution $\dst(\phi)$, we obtain the average
value of the observable $\tsth$ on the ensemble $\dst$, which reads
\begin{equation}\label{econstrain}
  \int_{\cbn}\;
  \tldh(\phi)
  \dst(\phi)\,
  \mesb
\;=\;
\left\langle
\tldh, \dst
\right\rangle
\end{equation}
where $\langle\cdot\,,\cdot\rangle$ is the scalar product in the
space $\mathcal{L}^2(\cbn,\mesb)$.

\medskip

On the other hand we may consider the average value $\trc\tsth\rrh$
of the observable $\tsth$ in state $\rrh$, and then consider only
the ensemble $\dst$ compatible with $\rrh$, in that sense that,
given $\tsth$ as the only measurement apparatus, we can not
distinguish $\rrh$ and any of such $\dst$. This is expressed as:
\begin{equation}\label{eaveragens}
\trc\tsth\rrh
\;=\;
\left\langle
\tldh, \dst
\right\rangle
\end{equation}

\section{The likelihood ratio}\label{s01}

The expression \eqref{eaveragens} specifies for us a class of
distributions. This is a broad class, for instance, it contains a
delta-like distribution
\(\dst(\phi)=\sum\lambda_j\,\delta(1-\braket{\phi}{\ee_j})\)
where $\sum \lambda_j\prjo{\ee_j}$ is the spectral decomposition of
the density matrix $\rrh$. Our basic suggestion is the following. We
fix certain \emph{fiducial} distribution on the set of pure
states---this distribution need not have anything in common with the
density matrix $\rrh$. In this paper the uniform distribution over
the set of all pure states is chosen as fiducial. Then, we compute a
distance between the distribution in question and fiducial one:
\(
\dst_{0}(\phi)
\;=\;
1
\),
which averages to completely mixed state
\begin{equation}\label{e1}
\int_{\cbn}\prjo{\phi}\mesb
\;=\;
\frac{\1}{\lth}
\;=\;
\left(
\begin{array}{cccc}
1/\lth &0 &\ldots&0\cr
0&1/\lth&\ldots&0\cr\ldots&\ldots&\ldots&\ldots\cr 0&0&\ldots&1/\lth
\end{array}
\right)
\end{equation}

There are many ways to define the distance between two
distributions; we specify it to be Kullback-Leibler distance
\cite{kullback}:
\begin{equation}\label{ekullbackgen}
\entrp(\dst\|\dst_0) \;=\;
\int \dst(x)\ln\frac{\dst(x)}{\dst_0(x)} dx
\end{equation}
The reason for this choice is that this distance minimizes the Type
I error when discriminating states. Let us dwell on this issue in
more detail.

Suppose the sender has two options: either emit states according
to the fiducial distribution, or emit them according to the
distribution $\dst$. This is a standard way to send classical
messages through quantum channels. The goal of the recipient is to
determine what happened on the sender's side. The recipient makes
a \emph{null hypothesis}---that the fiducial distribution was
applied. Another option will be referred to as the
\emph{concurring hypothesis}. The \emph{Type I error} is to accept
the concurring hypothesis in the case when the fiducial ensemble
was in fact prepared. The probability of Type I error is,
according to Sanov theorem
\[
p(1|0)
\;=\;
\rme^{-\entrp(\dst\|\dst_0)}
\]
That is why our goal is to minimize this probability, though staying
within the restriction \eqref{eaveragens}. In our setting we have a
particular choice \eqref{e1} for the fiducial distribution,
therefore the Kullback-Leibler distance will have the form
\begin{equation}\label{ekullback}
\entrp(\dst)\;=\;\int \dst(\phi) \ln \dst(\phi) \mesb
\end{equation}

\section{Conditional Gibbs distributions}

In this section we are going to solve a variational problem similar
to that arising in classical thermodynamics. For a given density
matrix $\rrh$ and given observable $\tsth$ we search a continuous
ensemble, denote its distribution by $\dst(\phi)$, having minimal
differential entropy with respect to the uniform ensemble and
satisfying the compatibility relation \eqref{eaveragens}:
\begin{equation}\label{egenprob}
\left\lbrace
\begin{array}{l}
\entrp(\dst)\;\to\; \min,
\cr
\trc\tsth\rrh
\;=\;
\left\langle
\tldh, \dst
\right\rangle
\end{array}
\right.
\end{equation}
---recall that $\langle\cdot\,,\cdot\rangle$ is the scalar product in the
space $\mathcal{L}^2(\cbn,\mesb)$. The combination of $\rrh$ and
$\tsth$ plays here the r\^ole similar to that of energy in classical
thermodynamics. We emphasize that the differential entropy used here
is a mixing entropy \cite{wehrl} (it is related to the ensemble
$\dst$) rather than von Neumann entropy, which depends only on the
density matrix $\rrh$. The ensemble which yields the solution of
this problem will be called \dfd{conditional Gibbs ensemble}. The
appropriate Lagrange function reads:
\[
\mathcal{L}(\dst) \;=\; \int \dst(\phi) \ln \dst(\phi)\mesb\;-\;
\lmu\left( \int \bracket{\phi}{\tsth}{\phi}\, \dst(\phi) \mesb -
\trc(\tsth\rrh) \right)
\]
where $\lmu$ is the Lagrange multiple. Making the derivative of
$\mathcal{L}$ over $\dst$ zero, we get
\begin{equation}\label{egibbs}
\dst(\phi) \;=\;
\rme^{-\tpar\bracket{\phi}{\tsth}{\phi}}\left/\vphantom{\int_S}Z\left(\tpar\right)\right.
\end{equation}
where $\tpar$ is the optimal value of the Lagrange multiple $\lmu$,
which we derive from the constraint \eqref{egenprob}. The
normalizing multiple
\begin{equation}\label{epartfun}
Z(\tpar) \;=\; \int \rme^{-\tpar\bracket{\phi}{\tsth}{\phi}} \mesb
\end{equation}
is the partition function for \eqref{egibbs}.

Yet having the explicit formula \eqref{egibbs} for the solution of
the variational problem \eqref{egenprob}, we still have to prove its
existence for any pair $\rrh$ and $\tsth$. To do that, first note
that that the differential entropy \eqref{ekullback} is a convcave
functional with respect to $\dst$. The restriction $\trc\tsth\rrh
\;=\; \left\langle \tldh, \dst \right\rangle$ is, in turn, linear,
thus specifying a linear affine manifold in the space
$\mathcal{L}^2(\cbn,\mesb)$. Then the existence of appropriate
$\dst$ directly follows from Gel'fand theorem \cite{gsr4}, and the
form of $\dst$ is specified by \eqref{egibbs}.

\section{Analytic expressions}

In this section we provide explicit formulas linking the temperature
parameter $\tpar$ with the density operator $\rrh$ and the measuring
observable $\tsth$. First let us evaluate the expression for the
average value of $\tsth$, that is, the lhs of \eqref{eaveragens}.
\[
  \int_{\cbn}\;
  \tldh(\phi)
  \dst(\phi)\,
  \mesb
\;=\;
\sum_{s=1}^{\lth}\limits\tsth_s
\pdr{\ln Z(\tpar)}{\thb_s}\;=\;
\gbs_{\tsth}(\tpar)
\]
The explicit expressions for the partial derivatives were obtained
earlier \cite{gogibbs}:
\[
\pdr{\ln Z(\tpar)}{\thb_s}
\;=\;-\frac{\frac{\rme^{-\tpar\thb_s}}{\prod_{j\neq{}s}\limits\thb_{sj}}
\;+\; \sum_{\stackrel{k=1}{k\neq{}s}}^{\lth}\limits
\frac{1}{\thb_{sk}} \cdot \left(
\frac{\rme^{-\tpar\thb_s}}{\prod_{j\neq{}s}\limits\thb_{sj}} +
\frac{\rme^{-\tpar\thb_k}}{\prod_{j\neq{}k}\limits\thb_{kj}} \right)
}{\sum_{k=1}^{\lth}\limits
\frac{\rme^{-\tpar\thb_k}}{\prod_{j\neq{}k}\limits\thb_{kj}}}
\]
where $\thb_s$ range over the eigenvalues of $\tsth$ and
$\thb_{sj}=\thb_s-\thb_j$ (if two or more of them are equal, the
appropriate expression is obtained as a limit starting with unequal
eigenvalues). Summing them up with the eigenvalues $\thb_s$ of
$\tsth$, we get the expression for $\gbs_{\tsth}(\tpar)$:
\begin{equation}\label{egrand}
\gbs_{\tsth}(\tpar)
\;=\;-\left(\sum_{s=1}^{\lth}\limits\,\thb_s\,
\frac{\rme^{-\tpar\thb_s}}{\prod_{j\neq{}s}\limits\thb_{sj}}\right)
\left/
\left(\sum_{s=1}^{\lth}\limits
\frac{\rme^{-\tpar\thb_s}}{\prod_{j\neq{}s}\limits\thb_{sj}}\right)
\right.
\end{equation}
From the expression \eqref{epartfun} for $Z(\tpar)$, we infer that
$\gbs$ is a monotonous decreasing function of the argument $\tpar$
whose values range between $+\infty$ and $0$; this takes place for
any $\tsth\neq 0$. That means, in turn, that the inverse of
$\gbs_{\tsth}$ exists for any $\tsth\neq 0$, denote it by
$\fbs_{\tsth}$:
\begin{equation}\label{efbs}
\fbs_{\tsth}
\;=\;
\gbs_{\tsth}^{-1}
\end{equation}
As a result, we may write down the formula for $\tpar$:
\begin{equation}\label{eexplbeta}
\tpar
\;=\;
\fbs_{\tsth}\left(\trc\,\tsth\rrh\right)
\end{equation}

\section{Analogs with temperature: equalizing and convexity}

Why do we claim that any observable $\tsth$ can be treated as
`thermometer'? Consider two quantum systems with state spaces $\hh$
and $\hh'$, respectively. Let their states initially be $\rrh$ and
$\rrh'$. Then, since we consider a non-interacting coupling of the
systems, the joint density matrix is
\(\prxp{\rrh}\) in the tensor product space \(\prxp{\hh}\). Let us
measure the sum of values of the observables $\tsth$ and $\tsth'$,
that is, introduce the observable
\(\tsthp=\tsth\otimes\1'+\1\otimes\tsth'\).

Now let us fix the fiducial distribution (introduced in section
\ref{s01}), in our case this will be the uniform distribution over
the set of \emph{product} pure states. This reflects the fact that
we are emitting particles in two independent laboratories. The
conditional optimal ensemble with respect to the observable
\(\tsthp\) is the following distribution
\[
\mu_{\tsthp}(\prxp{\psi}) \;=\; \rme^{-\beta_{\tsthp}\,
\bracket{\prxp{\psi}}{\tsthp}{\prxp{\psi}}}
\left/\vphantom{\int}Z_{\tsthp}\left(
\beta_{\tsth}\right)\vphantom{\int}\right.
\]
Like in classical thermodynamics, the partition function of the
joint system is the product of subsystems' partition functions:
\[
Z_{\tsthp}(\tau) \;=\; \iint
 \rme^{-\tau\,
\bracket{\prxp{\psi}}{\tsthp}{\prxp{\psi}}} \mesb\mesb' \;=\;\]\[=\;
\iint
 \rme^{-\tau\,
\left(\bracket{\psi}{\tsth}{\psi}+
\vphantom{\int}\bracket{\psi'}{\tsth'}{\psi'}\right)} \mesb\mesb'
\;=\; Z_{\tsth}(\tau) \cdot Z_{\tsth'}(\tau)
\]
therefore the equalizing property holds
\begin{equation}\label{eequrel}
\mbox{If}\quad \beta_{\tsth} \;\le\; \beta_{\tsth'}
\quad\mbox{then}\quad \beta_{\tsth} \;\le\; \beta_{\tsthp} \;\le\;
\beta_{\tsth'}
\end{equation}

\noindent which means that the conditional ensembles are equilibrium
and that $\beta$ plays the r\^ole of inverse temperature.

\medskip

As mentioned above, the function $\fbs_{\tsth}(\tpar)$, is a
monotone function of $\tpar$ when the observable $\tsth$ is fixed.
Let us explore $\fbs_{\tsth}\left(\trc\,\tsth\rrh\right) )$ as a
function of the density matrix $\rrh$, when the `thermometer'
$\tsth$ is fixed. From the convexity of $\fbs_{\tsth}(x)$ we deduce
that, when $\tsth$ is fixed, the function
$\fbs_{\tsth}\left(\trc\,\tsth\rrh\right) )$ is a convex function of
$\rrh$. That means, in turn, that the least value of the inverse
temperature $\tpar$ is attained on completely mixed state
\eqref{e1}. Conversely, the pure states are the `coldest' yielding
the maximum for the temperature parameter $\tpar$.

\section*{Conclusions}
Continuous ensembles of pure states proved their relevance in
various aspects of quantum mechanics. From the theoretical
perspective, they provide the limit cases on which numerical
characteristics of density matrices are attained, for instance, the
minimal value of accessible information about the state is attained
on `Scrooge' ensemble which is a continuous distribution
\cite{josubent}. Furthermore, I claim that they are relevant from
the operationalistic point of view. Even if we are speaking of
preparing discrete ensembles, we must also have in mind that their
are unavoidably smeared by various noises and, strictly speaking, we
have to deal with continuous distributions.

We introduce the notion of \emph{fiducial distribution} of pure
states, which can be seen as given `for free'. In our case this is
white noise---the uniform distribution on the set $\cbn$ of all pure
states of the system \cite{zaza}. Then we choose an observable
$\tsth$. For any given mixed state $\rrh$ a continuous ensemble is
shown to exist, which (i) has the smallest Kullback-leibler distance
from the fiducial one and (ii) reproduces $\rrh$ provided we can
measure only $\tsth$. The resulting ensemble is described by
exponential distribution \eqref{egibbs} of pure states:
\[
\left(1\left/\vphantom{\int}Z_{\tsth}\left(\tpar\right)\right.\right)\,
e^{-\tpar\bracket{\psi}{\tsth}{\psi}} \prjo{\phi}\mesb
\]
where the parameter $\tpar$ plays a r\^ole in some respect similar
to temperature, in particular, it is shown to possess the equalizing
property. The geometric properties of the proposed ensembles were
also studied \cite{0503173}.

The proposed ensembles can be applied for producing robust protocols
for sending classical messages through noisy quantum channels---this
is based on the general idea that in order to attain maximal
efficiency of communication, one must feed in mostly robust states
which are first of characterized by maximal entropy. Another
prospective application of the proposed techniques is quantification
of entanglement of mixed states \cite{0504169}.

\section*{Acknowledgments}

Profound discussion of the subject provided by prof. A.Kazakov and
the participants of St.Petersburg research seminar on Quantum
Information and Computation is acknowledged. The work was carried
out under the auspices of Russian Basic Research Foundation (grant
04-06-80215a). One of the authors (RRZ) highly appreciates the
hospitality of the Organizing Committee of Quantum-2006: III
workshop ad memoriam of Carlo Novero (in particular, Marco
Genovese) and the Quantum Computation Group of I.S.I. Foundation
(Torino, Italy) for valuable suggestions and comments within a
research seminar of the EC-funded project TOPQIP.



\vspace*{-5pt}   

\end{document}